\documentclass{article}

\usepackage[main,final]{neurips_2026}

\usepackage[T1]{fontenc}

\usepackage{amsmath, amssymb, amsthm}
\usepackage{thmtools}
\usepackage{amsfonts}
\usepackage{mathrsfs}
\usepackage[colorlinks=true,linkcolor=blue,citecolor=blue,urlcolor=blue,hypertexnames=false]{hyperref}
\hypersetup{%
  pdftitle={Fault Tolerant Coresets},
  pdfauthor={Milind Prabhu, Chris Schwiegelshohn, Sudarshan Shyam}
}
\usepackage{cleveref}
\usepackage{bm}
\usepackage{tcolorbox}

\usepackage{enumitem}
\usepackage{algorithm}
\usepackage{algorithmic}

\usepackage{tikz}
\usetikzlibrary{positioning,arrows.meta,shapes.geometric,calc}

\newtheorem{theorem}{Theorem}[section]
\newtheorem{lemma}[theorem]{Lemma}

\newtheorem{claim}[theorem]{Claim}

\theoremstyle{definition}
\newtheorem{definition}[theorem]{Definition}

\theoremstyle{remark}
\newtheorem{remark}[theorem]{Remark}

\title{Fault Tolerant Coresets}
\author{%
  Milind Prabhu \\
  University of Michigan \\
  Ann Arbor, MI, USA \\
  \texttt{milindpr@umich.edu}
  \And
  Chris Schwiegelshohn \\
  Aarhus University \\
  Aarhus, Denmark \\
  \texttt{schwiegelshohn@cs.au.dk}
  \And
  Sudarshan Shyam \\
  King's College London \\
  London, UK \\
  \texttt{sudarshan.shy@gmail.com}
}
\date{}

\newcommand{\eps}{\varepsilon}

\newcommand{\cost}{\textnormal{cost}}

\newcommand{\OPT}{\text{OPT}}

\usepackage{booktabs}

\newcommand{\T}{T}
\newcommand{\f}{f}

\usepackage{placeins}

\begin{document}
\maketitle

\begin{abstract}
Coresets are small weighted summaries of large datasets that help quickly
evaluate some loss objectives on the input dataset. We mainly focus on coresets
for the $k$-means objective, where the cost of the input dataset with respect to
a set of $k$ centers is the sum of squared distances from each point to its
closest center. A $k$-means $\eps$-coreset $\Omega$ satisfies
$\cost(P,S)=(1\pm\varepsilon)\cost(\Omega,S)$ for every set $S$ of $k$ centers,
where $P$ denotes the input dataset. A long line of work has established near
optimal $\eps$-coreset size bounds for $k$-means.  A key assumption in these works is that the coreset remains intact after construction. The question we ask is the following: if an adversary corrupts at most $f$ weighted entries of an $\eps$-coreset, can we recover an $O(\eps)$-coreset via a decoding algorithm that is given only the corrupted coreset. This models situations where entries of the coreset can be corrupted during storage or transmission, and recovery must proceed
without access to the original dataset. Our main result shows that increasing
the usual coreset size by an \emph{additive} $O(fk/\varepsilon)$ yields a coreset
that can be decoded into an $O(\varepsilon)$-coreset after any $f$ adversarial
corruptions. This bound is achieved by sensitivity sampling, a widely used
method for constructing coresets. Our result is optimal in the sense that any fault tolerant coreset, whose decoding algorithm only keeps a subset of the points
must also have size $\Omega(fk/\varepsilon)$.
Our experiments demonstrate recovery under adversarial corruptions.
\end{abstract}

\section{Introduction}
\label{sec:intro}
Modern datasets are often too large to store, process, or analyze directly.
Clustering is a basic tool for extracting structure from such data, and one of
the most widely used clustering objectives is $k$-means. Given a point set
$P \subseteq X$ in a metric space $(X,d)$, the goal is to choose a set
$S \subseteq X$ of $k$ centers minimizing
\[
    \cost(P,S)
    :=
    \sum_{p \in P} \min_{s \in S} d(p,s)^2 .
\]
Over the last two decades, \emph{coresets} have emerged as a successful tool for
addressing this challenge. At a high level, a coreset is a small weighted
summary of the input that approximates the objective value of every feasible
solution. Formally, given an input point set $P$, a weighted set $\Omega$ is an
$\eps$-coreset if
\[
    \cost(\Omega,S) \in (1\pm \eps)\cost(P,S)
    \qquad \text{for every set } S \text{ of } k \text{ centers}.
\]
Thus, one can run downstream clustering algorithms on $\Omega$ rather than on
the full input $P$, while preserving guarantees for the original instance. 
Following a long line of work, we now have very small coreset bounds for $k$-means in finite $n$-point metrics ($\tilde{O}(k \varepsilon^{-2}\log n)$ \cite{FL11}), metrics with bounded doubling dimension ($\tilde{O}(k\varepsilon^{-2}D)$ for doubling dimension $D$ \cite{CSS21}), and Euclidean
$k$-means ($\tilde{O}(k\varepsilon^{-2}\min(\sqrt{k},\varepsilon^{-2}))$ \cite{Cohen-AddadLSSS22,HLW23}). 
All of these bounds are optimal up to polylogarithmic factors.

These results assume that the coreset is accessed exactly as constructed. In many settings,
this assumption may be too strong. A coreset may be stored for long periods of
time, transmitted across unreliable channels, or passed through complex software
systems. In each case, some of its entries may be altered due to hardware faults,
packet loss, software errors, or adversarial attacks. This motivates the central question of this work.

\begin{tcolorbox}
Can coreset compression be made tolerant to adversarial
corruptions?
\end{tcolorbox}

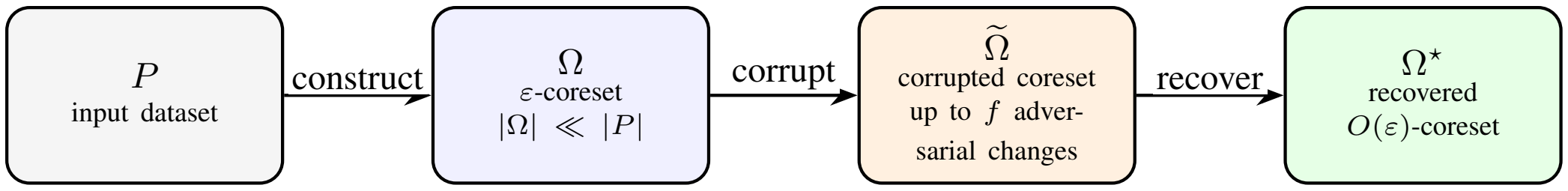
\begin{figure}[t]
\centering
\begin{tikzpicture}[
  font=\small,
  >=Stealth,
  node distance=1.3cm,
  arr/.style={-{Stealth[length=2.4mm]}, thick},
  lbl/.style={midway, above, font=\small, inner sep=1pt, fill=white},
  state/.style={
    draw,
    rounded corners=2mm,
    thick,
    minimum height=1.55cm,
    text width=2.2cm,
    align=center
  },
  data/.style={state, fill=gray!8},
  core/.style={state, fill=blue!6},
  corrupt/.style={state, fill=orange!12},
  recover/.style={state, fill=green!10}
]

\node[data] (P) {$P$\\[-0.5mm]\scriptsize input dataset};

\node[core, right=of P] (Omega)
{$\Omega$\\[-0.5mm]\scriptsize $\eps$-coreset\\[-0.5mm]\scriptsize $|\Omega|\ll |P|$};

\node[corrupt, right=of Omega] (Ot)
{$\widetilde{\Omega}$\\[-0.5mm]\scriptsize corrupted coreset\\[-0.5mm]\scriptsize up to $\f$ adversarial changes};

\node[recover, right=of Ot] (Os)
{$\Omega^\star$\\[-0.5mm]\scriptsize recovered\\[-0.5mm]\scriptsize $O(\eps)$-coreset};

\draw[arr] (P) -- node[lbl] {construct} (Omega);
\draw[arr] (Omega) -- node[lbl] {corrupt} (Ot);
\draw[arr] (Ot) -- node[lbl] {recover} (Os);

\end{tikzpicture}
\caption{Fault-tolerant coreset pipeline. From the input dataset $P$, we construct a coreset $\Omega$. An adversary then applies up to $\f$ corruptions, producing $\widetilde{\Omega}$. The goal is to recover from $\widetilde{\Omega}$ a set $\Omega^\star$ that is again a $O(\eps)$-coreset for the original dataset $P$.}
\label{fig:ft-coreset}
\end{figure}

We study this question in a post-compression corruption model. First, a coreset
$\Omega$ is constructed from the input $P$. Then an adversary may corrupt up
to $\f$ weighted entries of $\Omega$. Corrupting an entry means that both its
location and its weight may be changed arbitrarily. The decoder is given only
the corrupted summary $\widetilde{\Omega}$, not the original input $P$, and
must recover from it a valid coreset for $P$; see \Cref{fig:ft-coreset}.
Corruption models for samples drawn from a distribution have a long history in
robust statistics, beginning with the seminal works 
\cite{Tukey60,Huber64}, and continuing through later work on adversarial
contamination in  high-dimensional statistics (see \Cref{sec:related}).

The above question exposes a basic tension. The purpose of a coreset is to compress:
it achieves small size by discarding most of the redundant information in the input. Fault
tolerance, on the other hand, seems to require redundancy, since the summary
must remain useful even after some of its entries are corrupted. Thus, it is intriguing whether one can protect a coreset against corruptions without
giving up the compression that makes it useful.

\subsection{Our Contributions}
\label{suc:contributions}

We answer the main question of this paper in the affirmative for a broad class
of pointwise-loss objectives, including $k$-means. Our main conceptual
contribution is that sensitivity sampling, one of the most general and widely
used frameworks for coreset construction, can be made tolerant to
post-compression adversarial corruptions.

We first discuss the sensitivity sampling framework in its general form.
Formally, we consider objectives that decompose as sums of pointwise losses:
there is a solution space $\mathcal S$, and for every input point $p \in P$,
there is a nonnegative loss function $\ell_p : \mathcal S \to \mathbb R_{\ge 0}$
such that $L_P(S) = \sum_{p \in P} \ell_p(S)$ for every $S \in \mathcal S$.
The $k$-means objective is a central example: here $\mathcal S$ is the set of
all $k$-point subsets in the metric space in consideration, and
$\ell_p(S)=\min_{s \in S}d(p,s)^2$.

The sensitivity of a point $p \in P$ is
\[\sigma_P(p) := \sup\limits_{S \in \mathcal S \, : \,  L_P(S) > 0} \frac{\ell_p(S)}{L_P(S)},\] and the total
sensitivity is $\T := \sum_{p \in P} \sigma_P(p)$. Sensitivity sampling
constructs a coreset by sampling points with probability proportionate to upper
bounds on their sensitivities and reweighting the sampled points appropriately.
This framework is known to yield near-optimal coreset size bounds for a variety
of problems, including the aforementioned clustering results, but also linear regression \cite{Woodruff024a}, logistic regression \cite{MO24}, low-rank approximation
\cite{CohenMM17}, and spectral sparsification \cite{SpielmanS11}.

Our first result shows that the same framework can be made fault-tolerant. If a
standard sensitivity-sampling construction has total sensitivity $\T$ and
produces an $\eps$-coreset of size $m$, then adding
$O(\f \T/\eps)$ samples suffices to tolerate $\f$ adversarial corruptions of the
coreset.

\begin{theorem}[Fault-tolerant sensitivity sampling]
\label{thm:general-ft-sensitivity}
Fix $\eps \in (0,1/2)$ and $\f \in \mathbb N$. Consider any pointwise-loss
objective admitting a sensitivity-sampling $\eps$-coreset of size $m$ and total
sensitivity $\T$. Then there is an $\f$-fault-tolerant $O(\eps)$-coreset of size
\[
    m + O(\f \T/\eps).
\]
\end{theorem}

For a constant number of faults, the additive overhead is typically small
compared with standard sensitivity-sampling coreset sizes, which in many cases
are of order $\Omega(\T/\eps^2)$.
Nevertheless, one could ask if the $O(fT/\varepsilon)$ term could be improved.
It turns out that this is not possible, even for $k$-means.
Consider that for $k$-means, the total sensitivity is $\T=\Theta(k)$
\cite{LangbergS10} and thus, the additive overhead in
\Cref{thm:general-ft-sensitivity} becomes $O(\f k/\eps)$. 
The following theorem shows that this dependence on $\f$, $k$, and $\eps$ is
necessary in the worst case for coresets whose decoding algorithm only keeps a
subset of the points with their original weights.

\begin{theorem}[Lower bound for fault-tolerant $k$-means coresets]
\label{thm:ft-kmeans-lower-bound}
For every $\eps \in (0,1/2)$ and $\f \in \mathbb N$, there are $k$-means
instances for which any $\f$-fault-tolerant $\eps$-coreset supported on the input
points, whose decoding algorithm only keeps a subset of its entries with their
original weights, must have size $\Omega(\f k/\eps)$.
\end{theorem}

Fault tolerance is not only an existential question: ideally we would like an efficient decoding algorithm that can recover a valid coreset from the corrupted summary. In our
framework, the decoder relies on identifying points with unusually large
sensitivity. Consequently, the effectiveness of the decoder depends on
computing, or accurately approximating, sensitivities on the corrupted coreset.

While the sensitivities can be computed exactly up to numerical
precision for some problems,
this is a substantial challenge for $k$-means. Existing sensitivity-sampling implementations typically use
bicriteria solutions or other proxy upper bounds instead
\cite{FL11,BravermanFL16,sensitivitycoresets}. These upper bounds suffice for
sampling, but may substantially overestimate individual sensitivities and
therefore need not identify which points to remove. Here, we relate the
approximability of the $k$-means sensitivity of a point to the approximability of (constrained) $k$-means.

\begin{theorem}[Approximability of $k$-means sensitivities]
In finite metrics, pointwise sensitivity approximation for $k$-means is no
harder than constrained weighted $k$-means with admissible center sets, and is
at least as hard as finite-metric $k$-means up to arbitrarily small constant
slack.
\end{theorem}
In other words, we show that any $\alpha$-approximation algorithm can be used
to compute an $\alpha$-approximation to the sensitivity of a point $p$
(\Cref{thm:approx-ub}). Conversely, if it is NP-hard to approximate $k$-means
beyond a factor $\alpha$, then approximating the sensitivity beyond the factor $\alpha$ is likewise
NP-hard (\Cref{thm:sensitivity-hardness-main}). We review available algorithms in more detail in Appendix \ref{sec:sensitivity-hardness}.

\paragraph{Experiments.}
We conduct three sets of experiments. First, we test recovery under
Gaussian and adversarial corruptions. In an exact finite-domain
adversarial test on Adult's age coordinate ($k=2$, $m=1000$, $f=5$),
decoding reduces mean distortion from $46.875$ to $1.072$.
Second, on Taxi ($k=50$, $f=5$), sensitivity sampling achieves mean
witnessed distortion $1.124$, compared with $1.689$ for BICO
\cite{bico} and $1.686$ for StreamKM++ \cite{streamkm}.
Finally, on CoverType ($k=2$, $m=100$), sampling with exact
finite-domain sensitivities achieves mean empirical distortion
$1.080$, versus $1.115$ for the chosen proxy. A compression-regime
simplex example further separates these distributions even with
an optimal reference partition.

\subsection{Technical Overview}
\label{sec:technical-overview}

As a first observation, any coreset can be made tolerant to $\f$ corruptions by
storing $2\f+1$ labeled copies of each coreset entry and recovering each entry
by majority vote. This protects against $\f$ corruptions, but multiplies the
coreset size by $2\f+1$. Our goal is to avoid this multiplicative blow-up and
obtain fault tolerance with only an additive increase in size.

In what follows, we explain the main ideas using the $1$-means objective on the line. The same principle underlies the construction for more general cost functions. In $1$-means on the line, a solution
is a single center $c$, and the cost of a weighted set $Q$ is
\[
    \cost(Q,c)=\sum_{q\in Q} w(q) d(q,c)^2.
\]
A non-sampling coreset for $1$-means can be extremely compact: in fact, one can
construct a coreset with only two weighted points that preserves the mean and
variance \cite{FeldmanSS20}, and therefore preserves the cost of every center. Such a summary is
accurate when accessed exactly, but a corruption of one of its two weighted
entries can substantially change the objective it represents.

Coresets that concentrate substantial weight on a few representatives can be
vulnerable to corruption. Corrupting one high-weight summary point can
significantly change the represented objective. Our construction uses specific
properties of sensitivity sampling to avoid a multiplicative overhead in the
coreset size.

\paragraph{Why sensitivity sampling helps.}
Sensitivity sampling behaves differently: it spreads importance across sampled
points. Once the sampled points have sufficiently uniform importance, the cost
of every fixed solution is preserved by concentration
\cite{LangbergS10,FL11}. For $1$-means, after drawing a coreset of size $m$,
every uncorrupted sampled point has weighted sensitivity on the order of
$1/m$ with high probability; for $k$-means, the corresponding scale is
$O(k/m)$.

This pointwise control is the source of fault tolerance. Consider first a single
corruption. If the corrupted point has much larger sensitivity than the typical
scale, then it is detectable and can be removed. If its sensitivity is
comparable to the typical scale, then it may be impossible to identify, but its
effect on every candidate solution is already small. Thus, a corruption is
either detectable or harmless.

\paragraph{A Sensitivity-Based Iterative Decoder.}
For multiple corruptions, one might try to compute all sensitivities once and
remove the $\f$ largest-sensitivity points, an idea used in outlier-resistant coreset constructions
\cite{FL11,in2025robust} and in certain deterministic coreset constructions \cite{McCurdy18,PapailiopoulosKB14}. This fails because corruptions can mask one another. 
An example already exists for the $1$-means problem: Suppose the uncorrupted coreset of size $m$ lies at
the origin and the adversary corrupts two entries. They move one point $p_2$ to $1$ and another point $p_1$ to a very large positive value. For any solution, the contribution of $p_2$ to the cost will be strictly less than $\frac{1}{m-1}$, whereas the sensitivity of $p_1$ is close to $1$ and the sensitivity of the remaining $m-2$ points at the origin is at least $\frac{1}{m-1}$. A one-shot pruning rule according to sensitivities will always retain $p_2$ and the adversary successfully corrupts the coreset. 

Our decoder therefore uses an iterative peeling approach. Starting from the corrupted
coreset, it recomputes sensitivities with respect to the current weighted set
and removes a point of maximum sensitivity, repeating for $\f$ rounds. Notice how in the aforementioned example the sensitivity of $p_2$ immediately becomes $1$ after $p_1$ is removed.
It turns out this must be the case in general: At each step, either a high-influence corruption is exposed and removed, or all
remaining corruptions have small total effect on the cost of any solution. 
While the exposition here mainly focused on the $1$-means objective where sensitivities can be computed exactly, this is not the case for $k$-means. Nevertheless, the same arguments also work when using approximate sensitivities for each point.

\paragraph{Further comments and limitations.}
Our framework applies to problems that admit coresets via sensitivity sampling.
It does not immediately apply to settings where sensitivity sampling is not the
right coreset mechanism, such as certain fair clustering objectives \cite{SSS19}, or to
geometric approximation problems such as directional width
\cite{AHV04}. Whether fault-tolerant coreset constructions
with better than a multiplicative $O(\f)$ overhead exist for these problems
remains open.

\subsection{Related Work}
\label{sec:related}

\paragraph{Adversarial contamination and robust statistics.}
Our corruption model is conceptually related to adversarial contamination models
in robust statistics and learning, where a bounded fraction of samples may be
changed arbitrarily. Such models have a long history, beginning with the
classical works of \cite{Tukey60,Huber64}, and have recently
received significant attention in high-dimensional statistics and theoretical
machine learning \cite{DKKLT16,LRV16,BLMT22,CHLLN23,BV24}. This line of work
studies robust versions of basic statistical tasks, including mean and
covariance estimation, regression, stochastic optimization, and mixture
learning; see also the book of \citet{diakonikolas2023algorithmic} and the
survey of \citet{LugosiM19}.

\paragraph{Coresets.}
Coresets are a central tool for compressing large optimization instances while
approximately preserving the objective value of every feasible solution. The earliest examples focused primarily on geometric objectives \cite{AHV05}. The connection to learning theory was established in a series of papers \cite{Chen09,LangbergS10,FL11} and has been since studied for a variety of problems including clustering \cite{sensitivitycoresets,BecchettiBC0S19,BravermanCJKST022,braverman2021coresets,Cohen-AddadD0SS25,CGSS22,CSS23,schwiegelshohnsettling,HaM04,huang2018coresets,huang2020coresets}, regression \cite{BoutsidisDM13,DasguptaDHKM09,MaiMR21,MO24,MunteanuSSW18,Woodruff024a}, low-rank approximation
\cite{CohenMM17,FSS13,FengKW21,maalouf2020tight,SoW18,VaradarajanX12}, and spectral sparsification \cite{BansalST19,JambulapatiLS23,Lee018,Lee23,SpielmanS11}. Many of these results are based on sensitivity sampling, sometimes also known as leverage score sampling, or, in particular in the context of graph sparsification, sampling by electrical resistances. Indeed, for (variants of) these aforementioned problems, sensitivity sampling yields coresets that are optimal up to logarithmic factors. We refer to surveys for more pointers, see \cite{Feldman20}.

Somewhat related are robust coresets and coresets for outliers  \cite{FL11,huang2018coresets,coresetswithoutliers,HJW25,in2025robust}. Here, the cost of a query on the coreset after removing the most expensive points must remain the same. Nevertheless, as mentioned in the technical overview, there are differences and these methods do not seem to imply results for fault-tolerant coresets. 
The closest related work in spirit is the recent paper of
\citet{coresetswithnoise}, which studies coresets for clustering under stochastic noise.
There, the input dataset is corrupted by stochastic noise before the coreset is
constructed, and the coreset is built from the noisy input. In contrast, in our
model the coreset is first realized and then an adversary, with full knowledge
of the realized coreset, corrupts entries of the coreset itself.

\section{Model and Preliminaries}
\label{sec:preliminaries}

We use the following notation throughout. A pointwise-loss objective consists of
an input set $P$, a solution space $\mathcal S$, and nonnegative loss functions
$\{\ell_p:\mathcal S\to\mathbb R_{\ge 0}\}_{p\in P}$. For a solution
$S\in\mathcal S$, write $L_P(S):=\sum_{p\in P}\ell_p(S)$. For a weighted
multiset $Q$ with weights $w_Q$, we define $\ell_q^w(S):= \ell_q(S)\cdot w_q$ and analogously 
$L_Q^w(S):=\sum_{q\in Q}\ell_q^w(S)$. For $k$-means in a metric space
$(X,d)$, $\mathcal S$ is the collection of $k$-point subsets of $X$ and
$\ell_p(S):=d(p,S)^2:=\min_{s\in S}d(p,s)^2$.

A weighted multiset $\Omega$ is an $\eps$-coreset for $P$ if
$L_\Omega(S)\in(1\pm\eps)L_P(S)$ for every $S\in\mathcal S$.
Note that most coresets require using weighted losses.

\paragraph{The Corruption Model and Fault-Tolerant Coresets}
An adversary corrupts an entry $(q,w_\Omega(q))$ of a weighted multiset
$\Omega$ by replacing it with an arbitrary weighted entry $(q',w')$, where $q'$
lies in the ambient domain and $w'\ge 0$. For an integer $\f\ge 0$, an
$\f$-fault corruption of $\Omega$ is any weighted multiset
$\widetilde{\Omega}$ obtained by corrupting at most $\f$ entries of $\Omega$.

We now define what it means for a coreset to remain recoverable in this corruption model.

\begin{definition}[$\f$-fault-tolerant coreset]
Fix $\eps\in(0,1)$. A pair $(\Omega,\textsc{Dec})$ is an
\emph{$\f$-fault-tolerant $\eps$-coreset} for $P$ if, with constant probability
over the construction of $\Omega$, the following event holds:
\begin{enumerate}[label=(\roman*)]
    \item $\Omega$ is an $\eps$-coreset for $P$;
    \item for every $\f$-fault corruption $\widetilde{\Omega}$ of $\Omega$,
    the decoder returns a weighted submultiset
    $\Omega^\star=\textsc{Dec}(\widetilde{\Omega})\subseteq
    \widetilde{\Omega}$, with inherited weights, that is an $\eps$-coreset for
    $P$.
\end{enumerate}
\end{definition}

Conditioned on the success event for the initial coreset, the guarantee must
hold for every admissible adversarial corruption.

\section{Fault-Tolerant Coreset Framework}
\label{sec:framework}

In this section, we give a general fault-tolerant coreset construction for
objectives that admit sensitivity sampling. The initial summary is sampled by
standard sensitivity sampling: each point is sampled with probability
proportional to its sensitivity and assigned the corresponding
inverse-probability weight. The decoder uses approximate sensitivities only to
identify points to filter.

Let $\T$ denote the total sensitivity of $P$, and let
$m_{\mathrm{sc}}(\eps)$ denote the number of samples required for standard
sensitivity sampling to return an $\eps$-coreset for $P$ with constant
probability.

For a weighted multiset $Q$ with weights $w$, write
$\cost(Q,S):=\sum_{q\in Q}w(q)\ell_q(S)$. The sensitivity of an entry
$q\in Q$ is
\[
    \sigma_Q(q)
    :=\sup_{S\in\mathcal S:\,\cost(Q,S)>0}
    \frac{w(q)\ell_q(S)}{\cost(Q,S)}.
\]

For $\alpha\ge 1$, an $\alpha$-approximate sensitivity of a point $q$ in a
weighted multiset $Q$ is a value $\widetilde{\sigma}_Q(q)$ satisfying
\[
    \frac{1}{\alpha}\sigma_Q(q)
    \le
    \widetilde{\sigma}_Q(q)
    \le
    \sigma_Q(q).
\]Assume that, for some constant $\alpha\ge 1$, $\alpha$-approximate sensitivities can be computed efficiently for any point in any weighted set.

\begin{theorem}[Fault-tolerant coreset]
\label{thm:main-fault-tolerant}
For any $\eps\in(0,1/10)$ and $\f\in\mathbb N$, sensitivity sampling from $P$ with
\[
    m
    =
    m_{\mathrm{sc}}(\eps)
    +
    \frac{\alpha\f\T}{\eps}
\]
samples admits an efficient filtering decoder that recovers a $6\eps$-coreset
from any $\f$-fault corruption with constant probability over the construction
of $\Omega$.
\end{theorem}

The decoder is simple: it repeatedly recomputes approximate sensitivities on
the current corrupted summary and deletes a point of maximum approximate
sensitivity. The proof uses two facts. First, sensitivity sampling ensures that
every sampled copy has small weighted contribution
(\Cref{lem:smallpointcontribution}). Second, if the decoder ever deletes an
uncorrupted point, then all remaining corrupted points must have small
sensitivity and hence small total cost.

\begin{lemma}
\label{lem:smallpointcontribution}
Let $\Omega$ be obtained by sensitivity sampling $m$ points from $P$. Then, for
every sampled copy $x\in\Omega$ and every solution $S$,
\[
    w(x)\cost(x,S)
    \le
    \frac{\T}{m}\cost(P,S).
\]
\end{lemma}

\begin{proof}
Let $\sigma$ denote the sensitivities and let
$\T:=\sum_{p\in P}\sigma(p)$. Sensitivity sampling chooses point $p$ with
probability $q_p=\sigma(p)/\T$ and assigns weight $1/(mq_p)$.

For a sampled copy $x$, we have $w(x)=\T/(m\sigma(x))$. By the definition of
sensitivity,
\[
    \cost(x,S)\le \sigma(x)\cost(P,S)
\]
for every solution $S$. Hence
\[
    w(x)\cost(x,S)
    =
    \frac{\T}{m\sigma(x)}\cost(x,S)
    \le
    \frac{\T}{m}\cost(P,S). \qedhere.
\]
\end{proof}

\begin{algorithm}[H]
\caption{\textsc{Dec}$(\widetilde{\Omega},\f)$}
\label{alg:greedy-sensitivity-pruning}
\begin{algorithmic}[1]
\REQUIRE Corrupted weighted multiset $\widetilde{\Omega}$ and budget $\f$
\ENSURE Weighted multiset $\Omega^\star$
\STATE $Q_0\gets \widetilde{\Omega}$
\FOR{$i=1,\ldots,\f$}
    \STATE Compute an $\alpha$-approximation
    $\widetilde{\sigma}_{Q_{i-1}}(x)$ to $\sigma_{Q_{i-1}}(x)$ for each
    $x\in Q_{i-1}$
    \STATE Choose
    $x_i\in\arg\max_{x\in Q_{i-1}}\widetilde{\sigma}_{Q_{i-1}}(x)$
    \STATE $Q_i\gets Q_{i-1}\setminus\{x_i\}$
\ENDFOR
\STATE \textbf{return } $\Omega^\star\gets Q_\f$
\end{algorithmic}
\end{algorithm}

\begin{proof}[Proof of \Cref{thm:main-fault-tolerant}.]
Let $Q_i$ denote the weighted multiset remaining after $i$ deletions when
Algorithm~\ref{alg:greedy-sensitivity-pruning} is run on an arbitrary
$\f$-fault corruption $\widetilde{\Omega}$, so $Q_0=\widetilde{\Omega}$.
The algorithm outputs $Q_\f$. We will show that $Q_\f$ is a $6\eps$-coreset
for $P$. Condition on $\Omega$ being an $\eps$-coreset, an event of constant
probability, and write $Q_i=G_i\cup B_i$ for its uncorrupted and corrupted
entries.

\emph{Lower bound.} We show that
$\cost(Q_\f,S)\ge(1-6\eps)\cost(P,S)$ for every solution $S$. Each $G_i$
omits at most $\f$ corrupted samples and at most $i\le\f$ samples deleted by
the decoder. By \Cref{lem:smallpointcontribution} and
$m\ge\alpha\f\T/\eps$, the total lost cost is at most
$2\f\T\cost(P,S)/m\le2\eps\cost(P,S)$. Since $G_i\subseteq\Omega$,
we obtain, simultaneously for every solution $S$ and $0\le i\le\f$,
\begin{equation}
\label{eq:ft-uncorrupted-cost}
    (1-3\eps)\cost(P,S)
    \le \cost(G_i,S)
    \le (1+\eps)\cost(P,S).
\end{equation}
Since $Q_\f$ contains $G_\f$ and costs are nonnegative,
\[
    \cost(Q_\f,S)
    \ge \cost(G_\f,S)
    \ge (1-3\eps)\cost(P,S)
    \ge (1-6\eps)\cost(P,S).
\]

\emph{Upper bound.} We show that
$\cost(Q_\f,S)\le(1+6\eps)\cost(P,S)$ for every solution $S$.
If $B_\f=\emptyset$, then $Q_\f=G_\f$ and the bound
follows directly from \eqref{eq:ft-uncorrupted-cost}.
Otherwise, since the decoder makes $\f$ deletions and at most $\f$ entries
were corrupted, it must delete an uncorrupted point. Let $g$ be the first such
point, and let $Q=G\cup B$ be the summary just before its deletion. This
deletion lets us bound the sensitivity of every corruption still present.
By \Cref{lem:smallpointcontribution}, \eqref{eq:ft-uncorrupted-cost}, and
$Q\supseteq G$,
\[
    \sigma_Q(g)
    \le \frac{\T}{(1-3\eps)m}
    \le \frac{2\eps}{\alpha\f},
\]
where we used $\eps<1/10$. Since $g$ maximizes the approximate sensitivity,
every $b\in B$ satisfies
\[
    \sigma_Q(b)
    \le \alpha\widetilde{\sigma}_Q(b)
    \le \alpha\widetilde{\sigma}_Q(g)
    \le \alpha\sigma_Q(g)
    \le \frac{2\eps}{\f}.
\]
Summing over $|B|\le\f$ gives $\cost(B,S)\le2\eps\cost(Q,S)$ for every
$S$. Thus,
\[
    (1-2\eps)\cost(Q,S)
    \le \cost(G,S)
    \le (1+\eps)\cost(P,S).
\]
The remaining deletions can only decrease costs, so
\[
    \cost(Q_\f,S)
    \le \cost(Q,S)
    \le \frac{1+\eps}{1-2\eps}\cost(P,S)
    \le (1+6\eps)\cost(P,S),
\]
where the last inequality uses $\eps<1/10$. Together with the lower bound,
this proves that $Q_\f$ is a $6\eps$-coreset.
\end{proof}

\begin{remark}[Proxy sensitivity upper bounds]
\label{rem:proxy-sensitivities}
\Cref{thm:main-fault-tolerant} is stated for exact sensitivity sampling in the
initial construction. More generally, the same framework applies when the
initial sampling distribution is defined by proxy scores $\tau(p)$ satisfying
$\tau(p)\ge \sigma_P(p)$ for every $p\in P$.
Such sensitivity upper bounds are common in many coreset constructions, since they
are often easier to compute than exact sensitivities and can simplify the
analysis. See \Cref{app:proxy-sensitivities} for details.
\end{remark}

\paragraph{Tightness for $k$-means.}
The additive overhead in \Cref{thm:main-fault-tolerant} is tight for $k$-means.
The lower bound holds even for the simple Euclidean instance
$P=\{e_1,\dots,e_k\}\subset\mathbb R^k$ and for filtering decoders whose output
is a submultiset of the corrupted coreset with inherited weights.

\begin{theorem}[Lower bound for fault-tolerant $k$-means coresets]
\label{thm:lb}
Fix $\eps\in(0,1/2)$ and $\f\in\mathbb N$.
Let $P=\{e_1,\dots,e_k\}\subset\mathbb R^k$. Any
$\f$-fault-tolerant $\eps$-coreset $(\Omega,\textsc{Dec})$ for Euclidean
$k$-means on $P$, where $\Omega$ is a weighted multiset supported on $P$ and
$\textsc{Dec}$ only filters points with inherited weights, must satisfy
$|\Omega|=\Omega(\f k/\eps)$.
\end{theorem}

The proof uses the fact that any coreset for this instance must approximately
preserve the total weight placed on each basis vector. If too few copies of some
$e_i$ are stored, an adversary can zero out the largest $\f$ of them and remove
too much mass at $e_i$. A filtering decoder cannot recreate this mass, so the
decoded set fails to be a coreset. The full proof appears in
\Cref{app:ft-lower-bound}.
 \section{Computing Approximate Sensitivities for $k$-Means}
\label{sec:compute-sensitvities}

The decoder requires approximate pointwise sensitivities on weighted summaries.
We obtain these through a reduction to constrained weighted $k$-means with
admissible center sets. We focus on finite metrics, where any approximation
algorithm for this constrained problem yields sensitivity estimates with the
same approximation factor. For Euclidean instances, one may obtain the same
kind of oracle by first restricting centers to a sufficiently small candidate
set, incurring only a constant factor loss.

We next fix notation. Let $(X,d)$ be a finite metric space and let
$P\subseteq X$ be the input point set. We use the weighted cost, sensitivity,
and approximation conventions from \Cref{sec:framework}. For readability, we
present the argument for unweighted inputs; the weighted case follows by
replacing each point contribution by its weighted contribution.

We use the following constrained version of finite-metric $k$-means: given a
client set $P\subseteq X$ and a set $Y\subseteq X$ of admissible centers, the
goal is to choose $k$ centers $C\subseteq Y$ minimizing
$\sum_{p\in P}d(p,C)^2$.

\begin{theorem}[Finite Metrics]\label{thm:approx-ub}
Suppose there is a $\rho$-approximation algorithm for constrained
finite-metric $k$-means. Given a point $p\in P$, one can compute a
$\rho$-approximation to $\sigma_P(p)$ using $O(|X|)$ calls to this algorithm.
\end{theorem}

\begin{proof}
Fix $p\in P$, and let $S^\star$ be a set of centers maximizing the sensitivity
of $p$. Let $x^\star\in S^\star$ be the closest center to $p$, and write
$r^\star=d(p,x^\star)$. If we knew $r^\star$, then the best way to maximize the
ratio defining the sensitivity, among solutions whose closest center to $p$ is
at distance at least $r^\star$, would be to minimize the denominator. Thus the
problem reduces to a constrained $k$-means instance in which all centers must
lie outside the ball of radius $r^\star$ around $p$.

Of course, we do not know $r^\star$. Since centers are chosen from the finite
set $X$, the value of $r^\star$ is one of the distances from $p$ to a point of
$X$. We therefore try all such radii. For each radius $r$, we run the assumed
$\rho$-approximation algorithm with admissible centers restricted to points at
distance at least $r$ from $p$. Let $S_r$ be the returned center set, with cost
$\widetilde{\OPT}_p(r)=\cost(P,S_r)$, and define $\widetilde{\sigma}(p)$ to be the largest value
of $r^2/\widetilde{\OPT}_p(r)$ over all radii tried.

Every returned center set $S_r$ satisfies $d(p,S_r)\ge r$. Hence
\[
    \frac{r^2}{\widetilde{\OPT}_p(r)}
    \le \frac{d(p,S_r)^2}{\cost(P,S_r)}
    \le \sigma_P(p),
\]
so $\widetilde{\sigma}(p)\le\sigma_P(p)$. Conversely, $S^\star$ is feasible
for the constrained instance at radius $r^\star$, so the approximation
guarantee gives
$\widetilde{\OPT}_p(r^\star)\le\rho\cost(P,S^\star)$. Therefore,
\[
    \widetilde{\sigma}(p)
    \ge \frac{(r^\star)^2}{\widetilde{\OPT}_p(r^\star)}
    \ge \frac{(r^\star)^2}{\rho\cost(P,S^\star)}
    = \frac{\sigma_P(p)}{\rho}.
\]
Together, these inequalities show that $\widetilde{\sigma}(p)$ is a
$\rho$-approximation to $\sigma_P(p)$.

The algorithm tries at most $|X|$ radii, and therefore makes $O(|X|)$ calls to
the constrained $k$-means approximation algorithm.
\end{proof}
We complement the algorithmic reduction above with a matching hardness result.
The proof is deferred to \Cref{sec:sensitivity-hardness}.

\begin{theorem}[Hardness of approximating $k$-means sensitivities]
\label{thm:sensitivity-hardness-main}
Fix $\rho>1$. If $k$-Means is NP-hard to approximate within a factor $\rho$, then for every constant
$0<\delta<\rho-1$, it is NP-hard to approximate the $k$-means sensitivity
maximization problem in finite metrics within a factor $\rho-\delta$.
\end{theorem}

Together with the hardness of $k$-means
\cite{GuhaK99}, this implies that pointwise $k$-means sensitivity approximation
is NP-hard within every factor ${\gamma<1+8/e}$.

\section{Experiments}
\label{sec:experiments}

In this section, we give a detailed overview of our experimental results.
Codebase and results can be found \href{https://anonymous.4open.science/r/neurips26-1F0C/}{here}.
All experiments were run locally on an Apple-silicon CPU.

\paragraph{Datasets.} We use four standard datasets in our experiments: Adult \cite{adult_uci_1996}, MNIST \cite{mnist_uci_1998}, CoverType \cite{covertype_uci_1998}, and Taxi \cite{taxi_trajectory_uci_2013}; see \cref{table:datasets}.
In addition, we use synthetic datasets in some experiments to isolate specific failure modes of different coreset constructions.

\textbf{Evaluation of Coresets.} Coresets are difficult to evaluate, so we use the evaluation method initially proposed by \cite{SchwiegelshohnS22}. It consists of comparing the cost they have with respect to a diverse set of solutions with the corresponding cost on the original dataset.
The solution sets are constructed from $k$-means++-type solutions and problem-specific queries; in tractable cases, we evaluate exactly over the stated query domain.
For an evaluated solution set $\mathcal{Q}$, we report the distortion $\max_{S\in\mathcal{Q}}\max\left(\frac{\text{cost}(\Omega,S)}{\text{cost}(P,S)},\frac{\text{cost}(P,S)}{\text{cost}(\Omega,S)}\right)$, i.e. values close to \(1\) indicate that the coreset accurately preserves the $k$-means objective over the evaluated solutions.

\textbf{Experiments.}
Broadly speaking, we consider three experiments: (i) an evaluation of the distortion of fault tolerant sensitivity sampling coresets with random and adversarial faults, contrasting Gaussian perturbations motivated by \cite{coresetswithnoise} with our adversarial setting, (ii) a comparison between sensitivity sampling and two commonly used coreset implementations BICO by \cite{bico} and StreamKM++ by \cite{streamkm} for adversarial faults, and (iii) an empirical comparison of coreset quality using exact finite-domain sensitivities and practical sensitivity estimates, motivated by Section \ref{sec:compute-sensitvities}, against commonly used coarser sensitivity proxies such as those implemented and evaluated in \cite{schwiegelshohnsettling,SchwiegelshohnS22}. We focus on the second set of experiments here; additional details as well as a full description of the first and third set of experiments are in Appendix \ref{sec:appendix-experiments}.

\paragraph{Fault-tolerance of practical coreset algorithms.}
We compare the fault-tolerance of sensitivity sampling with that of BICO and StreamKM++. In the worst case, BICO and StreamKM++ are not fault tolerant, but worst case bounds are notoriously loose in practice.

For all algorithms, we perform an adversarial corruption by first computing a high quality solution $\mathcal{S}$ and then computing a coreset $\Omega$. We construct common queries by replacing one center of $\mathcal{S}$ at a time with a remote unused center, retaining exactly $k$ centers. For each query, the adversary sets to zero the weights of the $f$ coreset entries with largest weighted cost contributions. We report the maximum witnessed distortion over these query--corruption pairs. On $300{,}000$ Taxi endpoints, we use $k\in\{50,100\}$ and a coreset size cap of $2{,}000$, matching actual entry counts across algorithms in each trial.

\begin{figure}[h]
    \centering
    \includegraphics[width=0.95\linewidth]{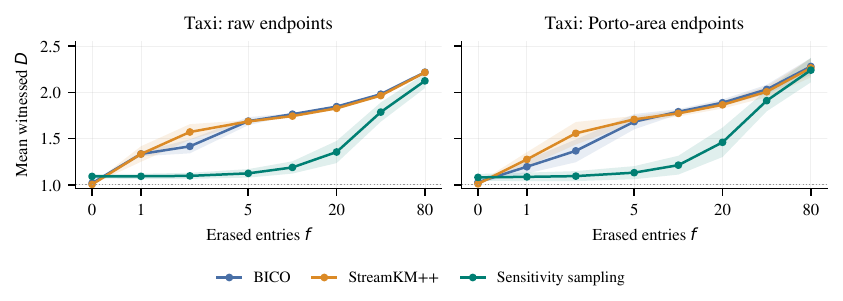}
    \caption{Comparison between BICO, StreamKM++ and sensitivity sampling on Taxi and geographically bounded Taxi endpoints, with $k=50$. Lines show means over $20$ trials; shading indicates one standard deviation.}
    \label{fig:comparison-allthree}
\end{figure}

Across our tested adversarial corruption settings, sensitivity sampling generally achieved the lowest mean witnessed distortion. On Taxi with $k=50$ and $f=5$, the mean distortion over $20$ trials was $1.124$ for sensitivity sampling, compared with $1.689$ for BICO and $1.686$ for StreamKM++. The corresponding clean distortions were $1.091$, $1.018$, and $1.004$. The advantage persisted on geographically bounded Taxi endpoints and narrowed at larger corruption budgets; see Figure \ref{fig:comparison-allthree}. On MNIST, BICO already had substantial clean distortion, so its post-corruption error does not isolate the effect of faults.

~

\begin{ack}
Milind Prabhu was supported by NSF grant CCF-2327011.
Chris Schwiegelshohn and Sudarshan Shyam were supported in part by a Google
Research Award. The authors declare no additional competing financial interests.
\end{ack}

\bibliography{ref}
\bibliographystyle{plainnat}

\newpage

\appendix

\section{Omitted Proofs from Section \ref{sec:framework}}
\subsection{Proxy Sensitivity Upper Bounds}
\label{app:proxy-sensitivities}
Typically, $k$-means coreset construction algorithms do not sample points proportional to exact sensitivities. Instead, points are sampled using proxy scores which are upper bounds on the sensitivity scores
\cite{FL11,FeldmanSS20,huang2020coresets,schwiegelshohnsettling,sensitivitycoresets}.  Their appeal
is twofold: they often lead to a cleaner analysis, and
the proxy scores are more easily computable.

For example a typical proxy score one might use is as follows. One can first compute a constant-factor
approximate solution $A=\{a_1,\ldots,a_k\}$ and let $C_j$ be the cluster
assigned to $a_j$. For each $p \in P$, 
\[
\tau_P(p)
    =
    \Theta\!\left(
        \frac{1}{|C_j|}
        +
        \frac{\cost(p,A)}{\cost(P,A)}
    \right)
    \qquad \text{for } p\in C_j.
\]

These scores are always upper bounds for the true sensitivities. However, they might be significantly overestimated. For example, in a high dimensional simplex, a good clustering may contain single points, and these will have their sensitivity overestimated by as much as a factor $n$.

We now show that our framework also applies to coreset constructions in which points are sampled proportional to such proxy scores. 

Let
$\tau:P\to\mathbb R_{\ge 0}$ satisfy $\tau(p)\ge \sigma_P(p)$ for every
$p\in P$, and let $\T_\tau:=\sum_{p\in P}\tau(p)$. Suppose the initial coreset
is constructed by sampling each point $p$ with probability
$q_p=\tau(p)/\T_\tau$ and assigning each sampled copy weight $1/(mq_p)$.

\begin{theorem}[Fault tolerance from proxy sensitivities]
\label{thm:proxy-sensitivity-ft}
Fix $\eps\in(0,1/10)$ and $\f\in\mathbb N$. Let
$m_{\mathrm{sc}}(\eps)$ be the sample size required by the above sampling
scheme to obtain an $\eps$-coreset for $P$ with constant probability. Assume
that $\alpha$-approximate sensitivities can be computed efficiently for any
point in any weighted set. If $\Omega$ is constructed with
\[
    m
    =
    m_{\mathrm{sc}}(\eps)
    +
    \frac{\alpha\f\T_\tau}{\eps}
\]
samples, then the filtering decoder of
\Cref{alg:greedy-sensitivity-pruning} outputs a $6\eps$-coreset for every
$\f$-fault corruption $\widetilde{\Omega}$ of $\Omega$, with constant
probability over the construction of $\Omega$.
\end{theorem}

\begin{proof}
The proof is identical to the proof of \Cref{thm:main-fault-tolerant}, with
\Cref{lem:smallpointcontribution} replaced by the following bound. For any
sampled copy $x$, its weight is $w(x)=\T_\tau/(m\tau(x))$. Since
$\tau(x)\ge \sigma_P(x)$, for every solution $S$,
\[
    w(x)\ell_x(S)
    =
    \frac{\T_\tau}{m\tau(x)}\ell_x(S)
    \le
    \frac{\T_\tau}{m\tau(x)}\sigma_P(x)L_P(S)
    \le
    \frac{\T_\tau}{m}L_P(S).
\]
Thus every occurrence of the bound $\T L_P(S)/m$ in the proof of
\Cref{thm:main-fault-tolerant} is replaced by
$\T_\tau L_P(S)/m$. The rest of the peeling analysis is unchanged, giving the
claimed guarantee.
\end{proof}

A natural question is whether these proxy upper bounds can also replace the
pointwise sensitivity approximations in the decoder. The following example
shows that a constant-factor approximation guarantee for the reference
clustering does not suffice: its proxy scores can strictly favor deleting a
genuine point over a harmful corruption.

\paragraph{Proxy scores are not enough for decoding.}
Fix $\eps\in(0,1/10)$ and an integer $n\ge8/\eps$. Consider $2$-means in
the shortest-path metric of the following graph. The input
$P=\{p_1,\ldots,p_n\}$ forms a complete graph with unit edge lengths.
Attach a vertex $a$ to $p_1$ by an edge of length $1/2$, and join a vertex
$b$ to every $p_i$ by an edge of length $M$, where $M^2=n-4$. Centers may
be any vertices. Take the exact clean summary $\Omega=P$, with unit weights.
Every two-center query assigns the same squared cost to all but at most two
input points, whose costs can only be smaller. Hence each input point has
sensitivity $1/(n-2)$, so the clean summary has uniformly small pointwise
contributions.

Corrupt $p_n$ by moving it to $b$, and call the resulting summary $Q$.
The reference centers $A=\{a,p_2\}$ induce a singleton cluster $\{p_1\}$
and a cluster containing all other points, including $b$. Their cost is
$D:=\cost(Q,A)=n-3+1/4+M^2=2n-27/4$. Any two-center solution costs at
least $n-3$ on the $n-1$ genuine points, so $A$ is a $2$-approximation.

Consider the usual proxy scores, scaled by a common factor:
\[
    \tau_Q(x)=2\left(\frac1{|C(x)|}+\frac{d(x,A)^2}{D}\right).
\]
The genuine singleton has score $\tau_Q(p_1)=2+1/(2D)>2$, whereas
\[
    1<\tau_Q(b)=\frac2{n-1}+\frac{2(n-4)}{D}<2,
\]
and every other score is also less than $2$. These scores are valid
sensitivity upper bounds: every genuine entry $q$ satisfies
$\sigma_Q(q)\le1/(n-3)\le2/(n-1)\le\tau_Q(q)$, by the same
common-cost observation, and $\sigma_Q(b)\le1<\tau_Q(b)$.

Thus the proxy-based decoder uniquely deletes $p_1$ and leaves the corruption.
For $S=\{p_2,p_3\}$, the decoded summary $\Omega^\star$ has distortion
\[
    \frac{\cost(\Omega^\star,S)}{\cost(P,S)}
    =\frac{2(n-4)}{n-2}
    \ge2-\eps>1+6\eps.
\]
In contrast, the same query shows that
$\sigma_Q(b)\ge(n-4)/(2n-7)>1/(n-3)$, so the exact-sensitivity decoder
uniquely deletes $b$. Its output is $P\setminus\{p_n\}$, an
$\eps$-coreset because removing one point loses at most a
$1/(n-2)\le\eps$ fraction of any query cost. The proxy rule therefore
fails even on an exact clean summary whose individual contributions are small.

\subsection{Lower Bound for Fault-Tolerant $k$-Means Coresets}
\label{app:ft-lower-bound}

In this section, we prove \Cref{thm:lb}. The construction is the simple
$k$-means instance $P=\{e_1,\dots,e_k\}\subset\mathbb R^k$. The proof shows that
a filtering decoder cannot recover from $\f$ corruptions unless the coreset
stores enough separate mass at each coordinate point.

For a weighted multiset $C$ supported on $P$, write
\[
    W_i(C):=\sum_{q\in C:\,q=e_i} w(q)
\]
for the total weight assigned to copies of $e_i$.

\begin{claim}[Weight preservation]
\label{clm:weight-pres}
If $C$ is an $\eps$-coreset for $P$, then
$W_i(C)\in[1-\eps,1+\eps]$ for all $i\in[k]$.
\end{claim}

\begin{proof}
Fix $i\in[k]$ and take
\[
    S_i:=\{e_1,\dots,e_{i-1},e_{i+1},\dots,e_k,0\}.
\]
Then $\cost(P,S_i)=\|e_i-0\|^2=1$. Moreover,
$\cost(C,S_i)=W_i(C)$, since only copies of $e_i$ contribute to the cost, each
with squared distance $1$. Since $C$ is an $\eps$-coreset for $P$,
\[
    W_i(C)=\cost(C,S_i)\in(1\pm\eps)\cost(P,S_i)=[1-\eps,1+\eps].
\]
\end{proof}

\begin{proof}[Proof of \Cref{thm:lb}]
Assume for contradiction that $|\Omega|<\f k/(2\eps)$. By pigeonhole, there is
some $i\in[k]$ such that $\Omega$ contains fewer than $\f/(2\eps)$ copies of
$e_i$. Let their weights be $a_1,\dots,a_m$, where
$m<\f/(2\eps)$, and let
\[
    A:=\sum_{j=1}^m a_j = W_i(\Omega).
\]
Since $\Omega$ is an $\eps$-coreset, \Cref{clm:weight-pres} gives
$A\in[1-\eps,1+\eps]$.

Let $J$ be the set of the $\min\{\f,m\}$ largest weights among
$a_1,\dots,a_m$. We claim that
\[
    \sum_{j\in J} a_j > 2\eps A.
\]
Indeed, if $m\le \f$, then $J=[m]$ and the sum is $A>2\eps A$ because
$\eps<1/2$. If $m>\f$, then the average weight among the $m$ copies is $A/m$,
so the sum of the $\f$ largest weights is at least
\[
    \f\cdot \frac{A}{m}
    >
    \f\cdot \frac{A}{\f/(2\eps)}
    =
    2\eps A.
\]

The adversary corrupts the entries indexed by $J$ by setting their weights to
$0$, obtaining an $\f$-fault corruption $\widetilde{\Omega}$. Let
$\Omega^\star:=\textsc{Dec}(\widetilde{\Omega})$. Since the decoder is a
filtering decoder, $\Omega^\star\subseteq\widetilde{\Omega}$ and inherits
weights. Therefore,
\[
    W_i(\Omega^\star)
    \le
    W_i(\widetilde{\Omega})
    =
    A-\sum_{j\in J} a_j
    <
    (1-2\eps)A
    \le
    (1-2\eps)(1+\eps)
    <
    1-\eps.
\]
By \Cref{clm:weight-pres}, $\Omega^\star$ cannot be an $\eps$-coreset for $P$.
This contradicts the definition of an $\f$-fault-tolerant $\eps$-coreset.
\end{proof}

\newpage
\section{Approximating Sensitivities: Omitted Proofs from Section \ref{sec:compute-sensitvities}}
\label{sec:sensitivity-hardness}

In this section, we prove \Cref{thm:sensitivity-hardness-main}, as well as give an overview of approximation algorithms and hardness of approximation and the implications on computing sensitivities. The reduction is
from the gap version of finite-metric $k$-means.

\begin{definition}[$\rho$-Gap $k$-Means Problem]
Given a finite-metric $k$-means instance and a threshold $B>0$, decide whether
the optimum $k$-means cost is at most $B$ or greater than $\rho B$, assuming one
of these is the case.
\end{definition}

It is known that, in general metrics, $k$-means is NP-hard to approximate
within a factor $1+8/e-\eta$ for every constant $\eta>0$
\cite{GuhaK99}. Equivalently, for every constant $\rho<1+8/e$, the
$\rho$-Gap $k$-Means problem is NP-hard.

\begin{proof}[Proof of \Cref{thm:sensitivity-hardness-main}]
Let $\alpha:=\rho-\delta$. Consider a finite-metric $k$-means instance given by
a point set $P$ in a metric space $(X,d)$. Let $\OPT$ denote its optimum
$k$-means cost, and let $D$ denote the diameter of $X$. We may assume $D>0$.

Construct a new metric space $X'=X\cup\{p^\star\}$ by adding a point
$p^\star$ with $d(p^\star,x)=D$ for every $x\in X$. Since $D$ is the diameter
of $X$, this defines a metric on $X'$. Assign weight $L$ to every point of
$P$, where $L$ will be chosen later, and add $p^\star$ with unit weight. Let
$P'$ be the resulting weighted instance.

We first relate the sensitivity of $p^\star$ in $P'$ to the optimum value of
the original $k$-means instance. We claim that
\[
    \sigma^*(p^\star)
    =
    \frac{D^2}{D^2+L\OPT}.
\]
Indeed, let $S$ be any set of $k$ centers in $X'$. If $p^\star\in S$, then
$\cost(p^\star,S)=0$, so the corresponding ratio is zero. Otherwise all centers
lie in $X$, and therefore $\cost(p^\star,S)=D^2$ and
\[
    \cost(P',S)=D^2+L\cost(P,S).
\]
Thus
\[
    \frac{\cost(p^\star,S)}{\cost(P',S)}
    =
    \frac{D^2}{D^2+L\cost(P,S)}.
\]
This ratio is maximized by minimizing $\cost(P,S)$, proving the claim.

Now suppose there is an $\alpha$-approximation algorithm for sensitivity
maximization. Define
\[
    A:=\frac{D^2}{D^2+LB}.
\]
If $\OPT\le B$, then $\sigma^*(p^\star)\ge A$, so the algorithm returns a
solution of value at least $A/\alpha$.

On the other hand, if $\OPT>\rho B$, then
\[
    \sigma^*(p^\star)
    <
    \frac{D^2}{D^2+L\rho B}.
\]
Choose $L$ large enough so that
\[
    \alpha(D^2+LB)<D^2+L\rho B.
\]
This is possible because $\alpha=\rho-\delta<\rho$. Hence
\[
    \sigma^*(p^\star)
    <
    \frac{D^2}{D^2+L\rho B}
    <
    \frac{D^2}{\alpha(D^2+LB)}
    =
    \frac{A}{\alpha}.
\]
Thus, by thresholding the algorithm's output at $A/\alpha$, we distinguish the
YES and NO cases of $\rho$-Gap $k$-Means. This contradicts the assumed
NP-hardness of the gap problem.
\end{proof}

\paragraph{Related Work on Approximation Algorithms}
As mentioned in the main body, the $k$-means objective cannot be approximated beyond a factor $1+8/e$ \cite{GuhaK99}. The problem remains hard to approximate in $\ell_p$ metrics including but not limited to Euclidean space, see \cite{AwasthiCKS15,Cohen-AddadS19,Cohen-AddadSL22,Cohen-AddadSL21,LeeSW17}. The approximation factor typically depends on the specific metric and, in some cases, on hardness conjectures stronger than P!=NP. Approximation algorithms for clustering in general metric spaces are an intensely studied topic \cite{AhmadianNSW20,ArthurV07,byrka2026k,Cohen-AddadEMN22,JainVazirani2001,KanungoMNPSW04}, with the current state of the art being a 5.83 approximation \cite{CharikarCGGLW25}. Further improvements are possible in constrained metrics such as low dimensional Euclidean spaces \cite{CKSS26}, planar graph metrics \cite{CKM16}, and doubling metrics \cite{Cohen-AddadFS21,FRS16}, or assumptions on the input \cite{AngelidakisMM17,ABS10,Cohen-AddadS17,ORSS12}.

\section{Further Experimental Evaluation}
\label{sec:appendix-experiments}

Here are the datasets used in the experiments.

\begin{table}[ht]
\centering
\begin{tabular}{l c}
\hline
Dataset & Shape after preprocessing \\
\hline
Adult & $(48842,\ 108)$ \\
MNIST & $(70000,\ 784)$ \\
CoverType & $(581012,\ 54)$ \\
Taxi & $(1704769,\ 2)$ \\
\hline
\end{tabular}
\caption{Datasets used with their shapes after preprocessing. Experimental subset sizes are specified below. Adult has 14 original features before encoding.}
\label{table:datasets}
\label{tab:datasets-shapes}
\end{table}

\paragraph{Preprocessing.}
For Adult, we exclude the income label, standardize the six numeric features, and one-hot encode the eight categorical features, retaining missing categories. The census sampling weight is treated as an ordinary numeric feature. For CoverType, we exclude the label and standardize all 54 features. For MNIST, we pool the training and test images, exclude labels, and divide pixel values by 255, without PCA or per-pixel standardization. For Taxi, we convert each nonempty trajectory into a two-dimensional point by taking the final GPS coordinate of the trip. The public download yields $1{,}704{,}769$ endpoints after excluding $5{,}901$ empty trajectories. We also use a geographically bounded control with longitude in $[-8.8,-8.2]$ and latitude in $[41.0,41.4]$, containing $1{,}701{,}157$ endpoints. Taxi subsets are translated and divided by a common scalar root-mean-square radius; this preserves clean and weight-erasure cost ratios. Full-feature experiments use ambient domain $\mathbb{R}^d$, and perturbation sizes below refer to the resulting coordinates.

All input subsets are sampled without replacement. Sampling coresets draw with replacement, assigning each sampled copy of point $i$ weight $1/(m p_i)$, where $p_i$ is its sampling probability. Copies remain separate entries. Decoder outputs retain inherited weights, without renormalization. Unless otherwise stated, results are means over the specified confirmation trials and error bars indicate one sample standard deviation; pilot runs are kept separately. Accuracy events use a numerical tolerance of $10^{-12}$.

\paragraph{Computational resources.}
All reported experiments were run locally on an Apple M1 CPU
(8 cores) with 16 GB of unified memory, running macOS 26.6.2
and Python 3.12.14. Numerical libraries were limited to two
threads per job; no GPU was used.
Approximate summed elapsed job times were 10 minutes for
fault recovery and its controls, 33 minutes for baseline
comparisons, and 35 minutes for sampling-distribution comparisons,
including the simplex controls, totalling approximately
1.3 job-hours.
The baseline total includes the interrupted-and-resumed
fixed-threshold control. Data preparation, environment setup,
additional pilot and validation runs, and postprocessing are
excluded; the full research required additional compute.

\subsection{Fault-Tolerant Coresets}

\paragraph{Sensitivity Sampling against random and adversarial faults.}
We first evaluate fault-tolerant coresets for $k$-means on the Adult, MNIST, CoverType and Taxi datasets. We construct a sensitivity-sampling coreset of size $m$ from the clean data and then corrupt a small number $f \ll m$ of its entries. We consider two corruption models: random corruption, where randomly selected coreset points receive Gaussian perturbations, and adversarial corruption, where selected points are moved far away in different directions and assigned much larger weights. Motivated by the stochastic-noise setting of \cite{coresetswithnoise}, we test Gaussian perturbations here after coreset construction, with our goal of recovering a good coreset.

For each coreset size $m$, we compare the clean coreset, the corrupted coreset, the fault-tolerant coreset, and a random-deletion control. The decoder repeatedly recomputes weighted sensitivity estimates on the current corrupted coreset and removes the most sensitive point for $f$ iterations. For the full-feature $k=10$ experiments, these estimates are exact only over a finite family generated from the corrupted summary: 128 randomly selected ten-point center sets and eight weighted or unweighted $k$-means solutions. This is a practical heuristic, without a certified approximation factor for unrestricted sensitivities. The decoder receives only the corrupted summary and $f$; clean data and corruption labels are used only for evaluation.

For downstream clustering, we run weighted $k$-means on each resulting coreset. All resulting center sets are evaluated on the original clean datasets, and we report the clean-data cost ratio relative to a full-data $k$-means reference solution. This reference is a high-quality local solution. For coreset quality evaluation, we use the evaluation approach of \cite{SchwiegelshohnS22} with a separate family of 32 clean-data queries: local $k$-means solutions, $k$-means++ seeds, random input-point center sets, and perturbed seeds. Coreset distortion and downstream clustering cost ratio are reported separately.

\begin{samepage}
\paragraph{Random corruptions.}
Let $\Omega=\{(x_i,w_i)\}_{i=1}^m$ be the coreset and let $\widetilde{\Omega}=\{(\widetilde{x}_i,\widetilde{w}_i)\}_{i=1}^m$ denote the corrupted coreset. We choose a corruption parameter $f\leq m$ and sample a subset uniformly at random:
\[
I\subseteq[m],\qquad |I|=f,
\]
\end{samepage}
For every $i\notin I$, we keep the point unchanged: $\widetilde{x}_i=x_i$. For every $i\in I$, we replace it by a randomly perturbed version. After the preprocessing above, all four datasets use Gaussian perturbations:
\[
\widetilde{x}_i=x_i+\sigma\xi_i,
\qquad \xi_i\sim\mathcal{N}(0,I_d),\qquad \sigma=0.2.
\]
All weights are unchanged. For MNIST, we use the same additive Gaussian pixel noise, without clipping. This model tests corruptions that need not dominate the clustering objective. In high dimension, the norm of a perturbation can nevertheless be appreciable.

\paragraph{Adversarial corruptions.}
Let $\Omega$ and $\widetilde{\Omega}$ denote the clean and corrupted coresets above. We choose a corruption parameter $f\leq m$ and select the $f$ entries with largest original weights, using a fixed rule to break ties. For every $i\notin I$, the point and weight remain unchanged. For every $i\in I$, we set
\[
\widetilde{x}_i=x_i+R u_i,\qquad
\widetilde{w}_i=\lambda w_i,
\]
where $R=80$, $\lambda=10$, and the $u_i$ are independent Gaussian directions normalized to unit length. This is an explicit admissible adversarial attack family, rather than an optimization over all possible faults. It checks whether the algorithm identifies and removes harmful points rather than allowing them to dominate the clustering objective.

\textbf{Experimental Setup.}
The main parameters are $k=10$ clusters, coreset sizes
\[
m\in\{100,200,500,1000\},
\]
and corruption level $f=0.05m$, giving $f\in\{5,10,25,50\}$ corrupted coreset entries. Each setting is averaged over 20 construction trials on a fixed 5,000-point subset per dataset. The initial ten-center coresets use the standard sensitivity proxy described below. The random-deletion control removes $f$ uniformly chosen entries from the same corrupted coreset. We plot coreset distortion as a function of $m$ and separately report the clean-data clustering cost ratios.

\textbf{Empirical Observations and Trends.}
The decoding algorithm recovers low empirical distortion under the specified adversarial corruptions. At $m=1000$ and $f=50$, mean decoded distortions are $1.088$, $1.160$, $1.076$, and $1.161$ on Adult, CoverType, MNIST, and Taxi, respectively, compared with corrupted distortions of approximately $1035$, $222.3$, $109.1$, and $66559$. The decoder removes all designated faults in these settings. The corresponding mean downstream cost ratios are $1.027$, $1.037$, $1.029$, and $1.116$.

\begin{figure}[!htbp]
    \centering
    \begin{minipage}{0.48\textwidth}
        \centering
        \includegraphics[width=\linewidth]{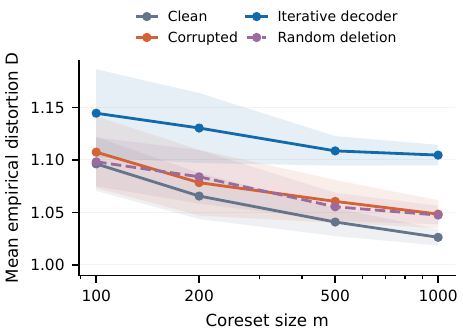}
        \caption{Distortion on Adult under random Gaussian corruptions. Means and one-standard-deviation bands over 20 trials; $k=10$, $f=0.05m$.}
        \label{fig:adult-random}
    \end{minipage}
    \hfill
    \begin{minipage}{0.48\textwidth}
        \centering
        \includegraphics[width=\linewidth]{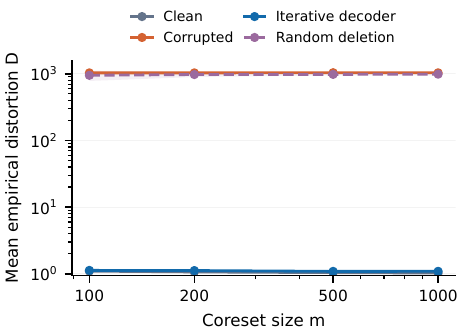}
        \caption{Distortion on Adult under adversarial coordinate and weight corruptions. Same settings as Figure \ref{fig:adult-random}; logarithmic vertical axis.}
        \label{fig:adult-adversarial}
    \end{minipage}
\end{figure}

\begin{figure}[!htbp]
    \centering
    \begin{minipage}{0.48\textwidth}
        \centering
        \includegraphics[width=\linewidth]{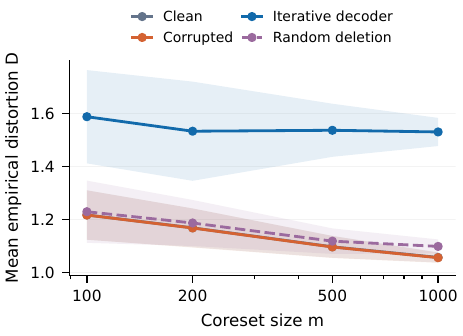}
        \caption{Distortion on CoverType under random Gaussian corruptions. Same settings as Figure \ref{fig:adult-random}.}
        \label{fig:covertype-random}
        \label{fig:covertype}
    \end{minipage}
    \hfill
    \begin{minipage}{0.48\textwidth}
        \centering
        \includegraphics[width=\linewidth]{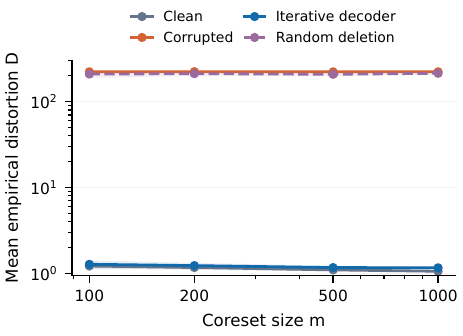}
        \caption{Distortion on CoverType under adversarial coordinate and weight corruptions. Same settings as Figure \ref{fig:adult-adversarial}.}
        \label{fig:covertype-adversarial}
    \end{minipage}
\end{figure}

\begin{figure}[!htbp]
    \centering
    \begin{minipage}{0.48\textwidth}
        \centering
        \includegraphics[width=\linewidth]{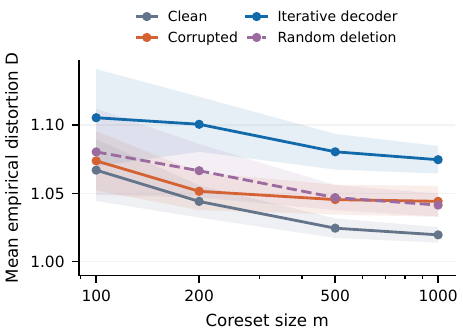}
        \caption{Distortion on MNIST under random Gaussian corruptions. Same settings as Figure \ref{fig:adult-random}.}
        \label{fig:mnist-random}
    \end{minipage}
    \hfill
    \begin{minipage}{0.48\textwidth}
        \centering
        \includegraphics[width=\linewidth]{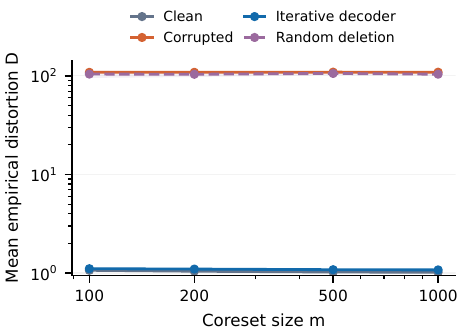}
        \caption{Distortion on MNIST under adversarial coordinate and weight corruptions. Same settings as Figure \ref{fig:adult-adversarial}.}
        \label{fig:mnist-adversarial}
    \end{minipage}
\end{figure}

\begin{figure}[!htbp]
    \centering
    \begin{minipage}{0.48\textwidth}
        \centering
        \includegraphics[width=\linewidth]{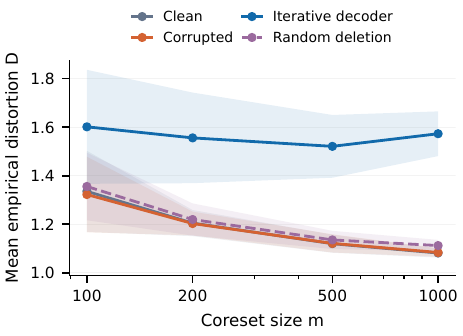}
        \caption{Distortion on Taxi under random Gaussian corruptions. Same settings as Figure \ref{fig:adult-random}.}
        \label{fig:taxi-random}
    \end{minipage}
    \hfill
    \begin{minipage}{0.48\textwidth}
        \centering
        \includegraphics[width=\linewidth]{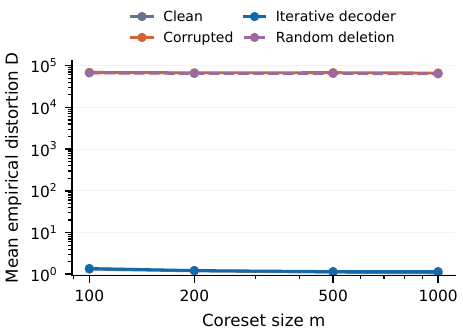}
        \caption{Distortion on Taxi under adversarial coordinate and weight corruptions. Same settings as Figure \ref{fig:adult-adversarial}.}
        \label{fig:taxi-adversarial}
    \end{minipage}
\end{figure}

Under random noise, decoding is often unnecessary and forced deletion can worsen coreset quality. At $m=1000$, the noisy distortions on CoverType and Taxi are $1.056$ and $1.084$, while decoding increases them to $1.530$ and $1.573$. Their decoded downstream cost ratios remain $1.018$ and $1.040$. The decoder makes all $f$ deletions even when the faults have little influence, and the discarded weight is not redistributed. Thus good downstream clustering does not imply that deletion preserves all evaluated query costs. Figures \ref{fig:adult-random}--\ref{fig:taxi-adversarial} show both regimes. Hence, our experiments show that the framework is effective for the specified adversarial corruptions, while mild noise may not require decoding.

\textbf{Exact recovery checks and additional controls.}
To distinguish the heuristic from exact decoding, we also run $k=1$ on the same four datasets, with 30 trials per setting. Weighted sensitivities have the closed form
\[
\sigma_Q(q_i)=\frac{w_i}{W}+\frac{w_i\|q_i-\mu_Q\|^2}{V_Q},\qquad
W=\sum_i w_i,\quad V_Q=\sum_i w_i\|q_i-\mu_Q\|^2,
\]
when $V_Q>0$. We recompute these sensitivities after every deletion and evaluate distortion analytically over all Euclidean single centers. At $m=1000$ and $f=50$, mean decoded distortions range from $1.101$ to $1.117$. In a separate fixed-budget check with $f=10$ and $m\in\{200,500,1000,2000\}$, decoded distortion decreases as $m$ grows; for Adult its mean decreases from $1.143$ to $1.028$.

For an exact two-center check, we use all 48,842 raw Adult ages and fix the public ambient domain
\[
X=\{-200,-100\}\cup\{0,1,\ldots,100\}\cup\{200,300\}.
\]
Every pair of distinct points in $X$ is a feasible query. The decoder receives this public domain, the corrupted summary, and $f$, but not clean ages or their frequencies. Sensitivities are recomputed exactly at every step, and evaluation covers all 5,460 center pairs. We use the same four sample sizes with $f=0.05m$, plus a separate $m=1000$, $f=5$ check, with 20 trials per setting. Adversarial faults replace the largest-weight entries by random elements of $\{-200,-100,200,300\}$ and multiply their weights by 10. Random faults add Gaussian age noise of standard deviation 2, then round and clip to $[0,100]$, keeping corruptions inside $X$.

At $m=1000$ and $f=5$, mean adversarial distortion falls from $46.875$ to $1.072$, compared with clean distortion $1.067$ and random-deletion distortion $46.138$. All five faults are removed in every trial. At $f=50$, the corresponding corrupted and decoded distortions are $445.311$ and $1.216$. Under random age noise, the noisy and decoded distortions are $1.067$ and $1.079$ at $f=5$, and $1.066$ and $1.254$ at $f=50$. The exact sensitivity total is $3.499$, giving the bound $fT/m=0.0175$ at $f=5$.

\begin{figure}[!htbp]
    \centering
    \begin{minipage}{0.48\textwidth}
        \centering
        \includegraphics[width=\linewidth]{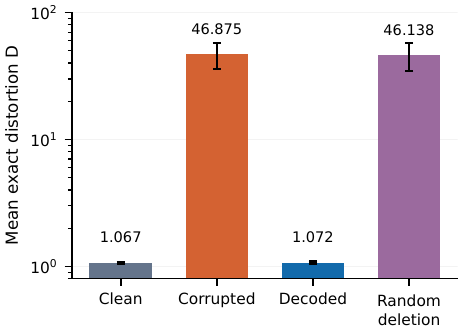}
        \caption{Exact Adult-age recovery, $k=2$, $m=1000$, $f=5$, over all 5,460 public-domain queries. Means and standard deviations over 20 trials.}
        \label{fig:adult-age-exact}
    \end{minipage}
    \hfill
    \begin{minipage}{0.48\textwidth}
        \centering
        \includegraphics[width=\linewidth]{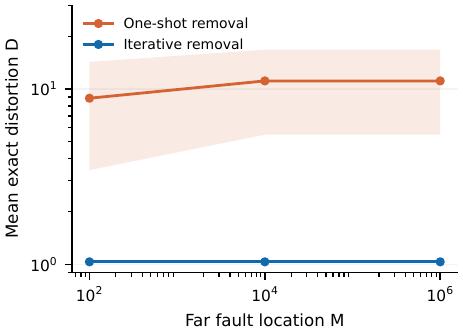}
        \caption{Exact one-center distortion under two masking faults. Means and standard deviations over 100 trials. Both axes are logarithmic.}
        \label{fig:masking-iterative}
    \end{minipage}
\end{figure}

Finally, we test the need to recompute sensitivities on a nondegenerate masking example: 5,000 one-dimensional $\mathcal{N}(0,0.01^2)$ points, $m=1000$, and two entries corrupted to locations 1 and $M\in\{100,10^4,10^6\}$. Over 100 trials, at $M=10^4$ one-shot removal gives mean exact distortion $11.112$ and removes only one fault, whereas iterative removal gives $1.034$ and removes both. Figure \ref{fig:masking-iterative} illustrates the difference.

\FloatBarrier
\subsection{Worst-Case Effect of Adversarial Corruptions on BICO and StreamKM++}

BICO and StreamKM++ both are \emph{movement-based} coresets. Such coresets use the property that a mapping $\pi$ of point set $P$ to another point set $Q$ results in an $O(\varepsilon)$-coreset if $\sum_{p\in P}d^2(p,\pi(p))\leq\varepsilon^2\cdot OPT$, where $OPT$ is the cost of an optimal $k$-means clustering. This mapping is typically used to summarize a large number of very close points by choosing the image to be the same point $q$ and weighting $q$ by the number of points in the pre-image of $q$.
This property alone does not imply fault tolerance. For $k=1$, consider a point set $P$ consisting of $n$ colocated points and a coreset $\Omega$ consisting of one of these points weighted by $n$, which would also be the summary generated by BICO and StreamKM++ on this instance. Moving the sole point in $\Omega$ to a different location corrupts the coreset with no possibility of recovery by filtering with inherited weights.

The goal of the experiment is to test whether adversarial corruption can cause cost underestimation that a filtering decoder cannot repair. Given a candidate clustering solution $S$, the adversary sets to zero the weights of the $f$ coreset entries with largest weighted cost to $S$. This changes at most $f$ entries and is admissible in our model, which permits arbitrary nonnegative replacement weights. Coordinates and all other weights remain unchanged, and the original dataset $P$ is not modified.

Specifically, let $I_f(S)$ index the $f$ largest values $w_i d(x_i,S)^2$ and let $\widetilde{\Omega}_{f,S}$ denote the resulting corrupted coreset. Its cost is
\[
C_f(S)=\operatorname{cost}(\Omega,S)-\sum_{i\in I_f(S)}w_i d(x_i,S)^2.
\]
For a common query family $\mathcal{Q}$, the quantity measured is the maximum witnessed distortion
\[
D_f^{\mathrm{wit}}=
\max_{S\in\mathcal{Q}}\max\left\{
\frac{C_f(S)}{\operatorname{cost}(P,S)},
\frac{\operatorname{cost}(P,S)}{C_f(S)}\right\}.
\]
Each query can induce a different corruption; this is a maximum over query--corruption pairs, rather than evaluation of one common corrupted coreset on every query. Whenever $C_f(S)<\operatorname{cost}(P,S)$, every weighted submultiset $Q\subseteq\widetilde{\Omega}_{f,S}$ with inherited weights satisfies
\[
\operatorname{cost}(Q,S)\leq C_f(S).
\]
Consequently, the reciprocal ratio for that witness is a lower bound on the distortion of every permitted filtering decoder. We do not apply a solution-aware trimming decoder in this comparison. Sensitivities themselves are well-defined for movement-based summaries; the sampling-based recovery theorem does not automatically apply to their construction.

\textbf{Experiment Details.}
We ran this test on the Taxi and MNIST datasets. For Taxi, we used a random subset of 300,000 endpoints, with dimension $d=2$, and repeated the analysis on 300,000 geographically bounded endpoints. For MNIST, we used 10,000 images represented as vectors in $\mathbb{R}^{784}$. We tested $k\in\{50,100\}$ and corruption budgets
\[
f\in\{0,1,2,5,10,20,40,80\}.
\]
Each case uses 20 construction trials on a fixed input subset. The primary BICO implementation follows the published insertion and rebuilding procedures with threshold doubling \cite{bico}. StreamKM++ uses weighted TreeCoreset splitting and binary merge-and-reduce \cite{streamkm}. These are fresh implementations from the published descriptions; optional nearest-neighbor accelerators are omitted. Sensitivity sampling uses the standard proxy below.

For each dataset and value of $k$, we first computed a high-quality reference clustering $S_{\mathrm{ref}}$ on the original dataset. The common query bank consists of $k$ queries: each removes one reference center and inserts a remote unused center, retaining exactly $k$ centers. It is fixed independently of the coreset draws and used for all methods. The adversary then chooses the largest weighted contributors separately for each query and method.

We use a 2,000-entry cap on Taxi and a 1,000-entry cap on MNIST. BICO's actual entry count is matched exactly by StreamKM++ and sensitivity sampling in each trial. Actual counts range from 1,754 to 1,850 on raw Taxi, 1,231 to 1,999 on geographically bounded Taxi, and 408 to 987 on MNIST. The $f=0$ results are the clean controls. We report each corruption budget separately rather than averaging over different values of $f$.

\textbf{Empirical Observations and Trends.}
Taxi provides a clear fault-specific separation. With $k=50$ and $f=5$, mean witnessed distortions are $1.689$ for BICO, $1.686$ for StreamKM++, and $1.124$ for sensitivity sampling, compared with clean values $1.018$, $1.004$, and $1.091$. With $k=100$, the corresponding attacked distortions are $1.817$, $1.770$, and $1.299$. The separation survives geographic restriction: for $k=50$ and $f=5$, mean distortions are $1.683$, $1.708$, and $1.132$. This is consistent with concentrated movement summaries losing important represented mass when a few entries are erased. The sampling advantage narrows substantially at larger budgets; for example, at $k=50$ and $f=80$ on raw Taxi, the means are $2.218$, $2.216$, and $2.125$.

\begin{figure}[!htbp]
    \centering
    \includegraphics[width=\linewidth]{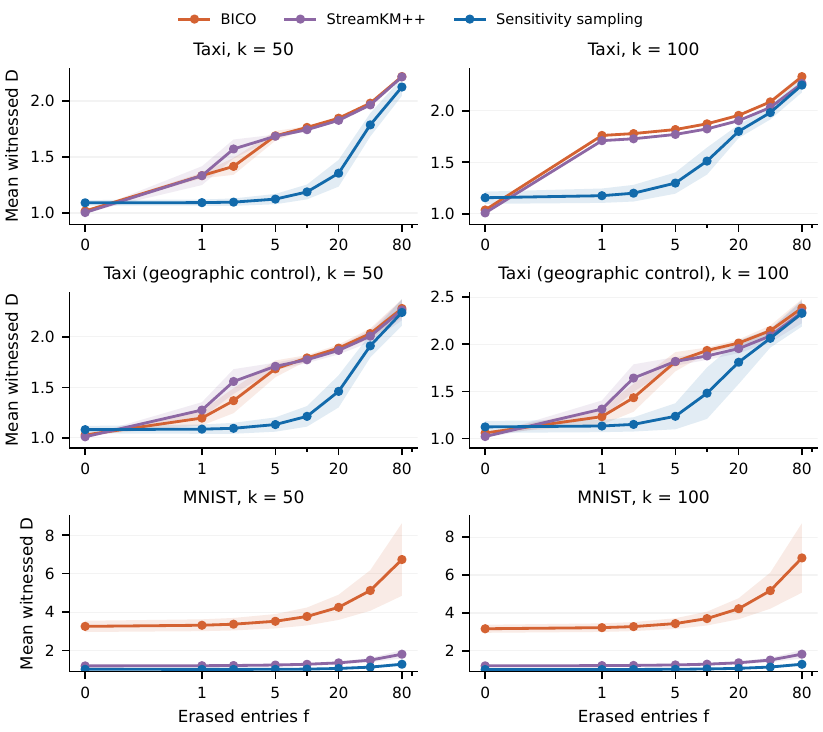}
    \caption{Comparison between BICO with rebuilding, StreamKM++ and sensitivity sampling under targeted weight erasures. Rows show raw Taxi, geographically bounded Taxi, and MNIST; columns show $k=50$ and $k=100$. Lines show means over 20 trials, shading indicates one standard deviation, and $f=0$ is the clean control. Entry counts are matched within each trial.}
    \label{fig:baseline-erasure-appendix}
\end{figure}

MNIST is more sensitive to the clean approximation quality of the implementation. At $k=50$, BICO's clean distortion is already $3.258$, compared with $1.201$ for StreamKM++ and $1.023$ for sensitivity sampling; at $f=20$, these become $4.249$, $1.359$, and $1.069$. Thus BICO's large post-corruption MNIST error does not by itself isolate fault damage. A separate fixed-threshold BICO control, with entry counts again matched, gives the same qualitative Taxi separation; for $k=50$ and $f=5$, the means are $1.665$, $1.669$, and $1.116$. Figure \ref{fig:baseline-erasure-appendix} shows the primary full-budget comparison.

\FloatBarrier
\subsection{Point-Wise Approximated Sensitivities vs Sensitivity Proxies}

Third, we evaluate the practical implications of computing sensitivities accurately, which is also one of our contributions in Section~\ref{sec:compute-sensitvities}. Many existing implementations rely on proxy upper bounds rather than computing sensitivities directly. We therefore compare coresets constructed using standard proxy sensitivities with coresets constructed using exact finite-domain sensitivities or practical sensitivity estimates. We perform this comparison on standard datasets as well as on a synthetic instance designed to highlight failure modes of the proxy sensitivity distribution.

As mentioned in Section \ref{app:proxy-sensitivities}, the commonly understood algorithm referred to as sensitivity sampling uses slightly different probabilities than point-wise approximate sensitivities. The most powerful of these proxies also have optimal worst-case guarantees \cite{sensitivitycoresets,schwiegelshohnsettling} and are empirically competitive among previous coreset implementations \cite{SchwiegelshohnS22}. Here, we investigate whether more accurate sensitivities offer an empirical improvement over a standard proxy, complementing the analysis of \cite{SchwiegelshohnS22}. Our comparison concerns the proxy specified below, rather than every optimal or fast proxy variant.

This experiment compares four sampling distributions for $k$-means coresets on Adult, MNIST, CoverType, and Taxi: uniform sampling, lightweight coreset sampling \cite{BachemL018}, the standard proxy sensitivity, and exact or estimated sensitivities. The proxy is proportional to
\[
\tau_i=\frac{1}{|C(i)|}+\frac{\|x_i-c(i)\|^2}{\sum_j\|x_j-c(j)\|^2},
\]
where $c(i)$ and $C(i)$ are the assigned reference center and cluster. Lightweight sampling mixes uniform probabilities with squared distance to the global mean in equal proportions. All scores are normalized to sampling probabilities. Each coreset is evaluated over the same fixed candidate center family for that dataset and setting. The goal is to see whether the sensitivity scores give lower distortion, lower variance, or smaller required coreset size than the usual proxy while keeping evaluation identical across methods.

\textbf{Experimental Setup.}
For the practical ten-center experiment, the dataset is restricted to $n_{\mathrm{exp}}=1000$ points, with $k=10$, and coreset sizes
\[
m\in\{100,200,500,1000\},\qquad m/k\in\{10,20,50,100\}.
\]
We use three independently sampled subsets per dataset and 200 draws per subset and size. The candidate center domain is 64 input points selected by $D^2$ sampling. For each input point, five radii restrict the admissible centers while leaving at least ten candidates; greedy center addition approximately solves each restricted problem. The resulting scores are feasible-solution lower estimates, not certified constant-factor approximations. Evaluation uses 128 independently generated ten-center subsets of the same domain. We plot mean distortion with standard-deviation bands and the 90th percentile of distortion against $m$ on a logarithmic horizontal axis.

For exact scores, we use five independently sampled 1,000-point subsets per dataset and $k=2$. The center domain is the set of distinct input locations, and we enumerate every pair of distinct candidate centers to compute sensitivities. Evaluation uses a separate family of 128 pairs, drawn from snapped local solutions, $k$-means++ seeds, random pairs, and distant pairs. Each setting uses 200 draws per subset and size. In an exhaustive audit, we use three fresh 200-point subsets per dataset, $m\in\{25,50,100\}$, and 250 draws per subset, evaluating both sensitivities and distortion over every one of the 19,900 center pairs. Exactness in these tests refers to the specified finite domain, not arbitrary Euclidean centers.

\textbf{Empirical Observations and Trends.}
On real-world datasets, accurate scores improve over the chosen proxy in specific tested settings. On CoverType at $m=100$, the practical $k=10$ scores give mean empirical distortion $1.152$, versus $1.274$ for the proxy and $1.167$ for lightweight sampling. The exact finite-domain $k=2$ comparison gives $1.080$ versus $1.115$ for the proxy; lightweight sampling is essentially tied at $1.080$. The proxy-minus-exact difference is positive on all five tested subsets, with mean $0.0355$ and a conditional paired 95\% bootstrap interval $[0.0325,0.0385]$. This interval describes sampling variation within the tested subsets, not population-wide uncertainty.

\begin{figure}[!htbp]
    \centering
    \includegraphics[width=\linewidth]{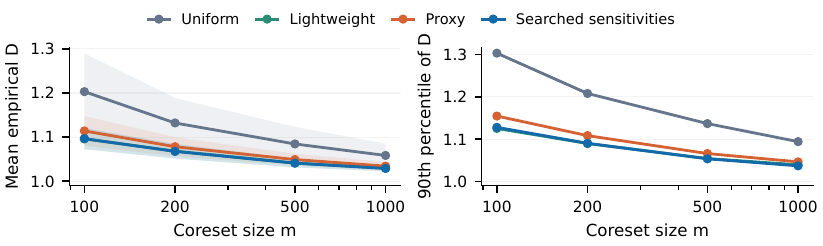}
    \caption{Comparison of sampling distributions on Adult, $k=10$, with a 64-point candidate center domain. Left: mean empirical distortion with standard-deviation bands. Right: 90th percentile. Three subsets and 200 draws per subset; the searched sensitivities are heuristic estimates.}
    \label{fig:adult-distortion}
\end{figure}

\begin{figure}[!htbp]
    \centering
    \includegraphics[width=\linewidth]{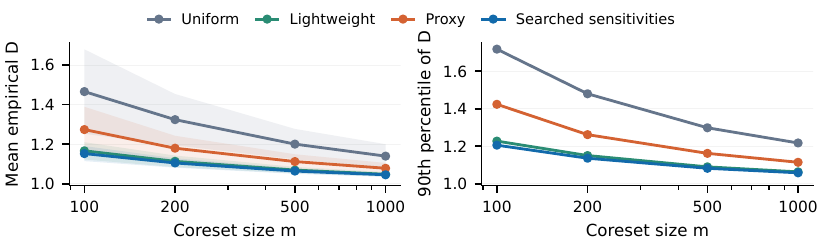}
    \caption{Comparison of sampling distributions on CoverType. Same settings and quantities as Figure \ref{fig:adult-distortion}.}
    \label{fig:covertype:distortion}
\end{figure}

\begin{figure}[!htbp]
    \centering
    \includegraphics[width=\linewidth]{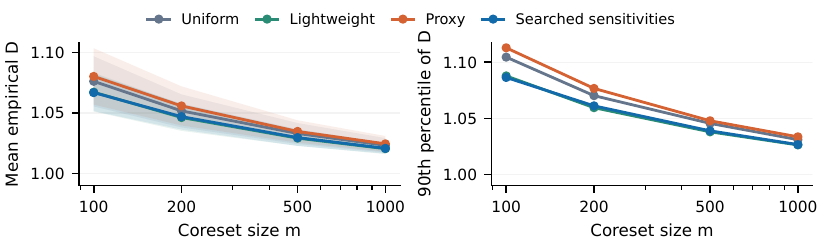}
    \caption{Comparison of sampling distributions on MNIST. Same settings and quantities as Figure \ref{fig:adult-distortion}.}
    \label{fig:mnist:distortion}
\end{figure}

\begin{figure}[!htbp]
    \centering
    \includegraphics[width=\linewidth]{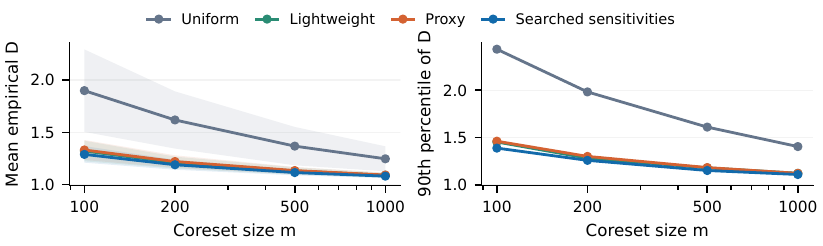}
    \caption{Comparison of sampling distributions on Taxi. Same settings and quantities as Figure \ref{fig:adult-distortion}.}
    \label{fig:taxi-distortion}
\end{figure}

\begin{figure}[!htbp]
    \centering
    \includegraphics[width=\linewidth]{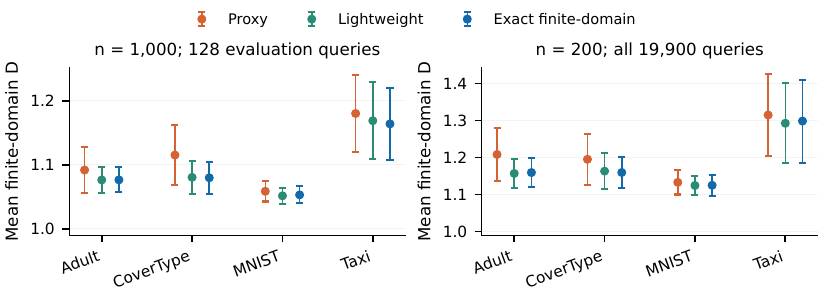}
    \caption{Exact finite-domain sensitivity comparisons. Left: $n=1000$, $m=100$, five subsets and 200 draws per subset, evaluated on 128 queries. Right: $n=200$, $m=50$, three subsets and 250 draws per subset, evaluated over all 19,900 queries. All queries have $k=2$ distinct centers. Points show means; error bars show one standard deviation.}
    \label{fig:exact-score-audits}
\end{figure}

The exhaustive 200-point audit retains favorable gaps: at $m=50$, Adult gives mean distortion $1.159$ for exact scores versus $1.208$ for the proxy, and CoverType gives $1.160$ versus $1.195$. Lightweight sampling again performs comparably. Figures \ref{fig:adult-distortion}, \ref{fig:covertype:distortion}, \ref{fig:mnist:distortion}, and \ref{fig:taxi-distortion} show the practical ten-center results; Figure \ref{fig:exact-score-audits} gives the exact finite-domain comparisons. These experiments support specific improvements over the chosen proxy without establishing universal superiority over other samplers.

\FloatBarrier
\textbf{Description of the Synthetic Dataset.}
The simplex experiment is designed to expose the waste in the proxy sensitivity term $1/|C|$.

We take $n$ vertices of a regular simplex, represented by the unit basis vectors $x_i=e_i$, so by symmetry the true sensitivity distribution is uniform. We fix a reference partition that makes the first $k-1$ vertices singleton clusters and puts all remaining vertices into one large cluster. This is an optimal $k$-means partition: a cluster of $s$ vertices has centroid cost $s-1$, so every nonempty $k$-partition has total cost $n-k$. Thus the example does not require a bad bicriteria fit.

Writing $r=n-k+1$, the chosen proxy gives each singleton score 1 and each point in the large cluster score $2/r$. It therefore assigns total probability $2/(k+1)$ to the large cluster. To reveal the performance gap, we first choose $k$ comparable to $n$, namely $n=300$ and $k=200$. The proxy spends most of its mass on the $k-1$ singleton vertices even though they are not more sensitive. Uniform and lightweight sampling coincide with normalized exact sensitivities on this symmetric instance.

\begin{figure}[!htbp]
    \centering
    \begin{minipage}{0.48\textwidth}
        \centering
        \includegraphics[width=\linewidth]{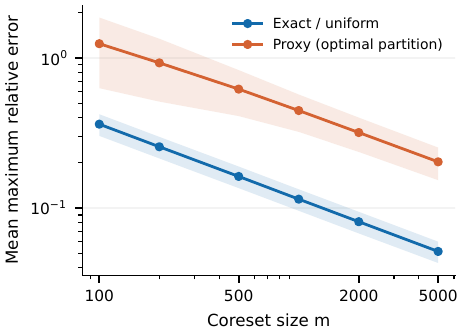}
        \caption{Mean maximum relative cost error on the small simplex, over 64 fixed queries and 10,000 draws per method and size. Bands show one standard deviation.}
        \label{fig:mean-simplex}
    \end{minipage}
    \hfill
    \begin{minipage}{0.48\textwidth}
        \centering
        \includegraphics[width=\linewidth]{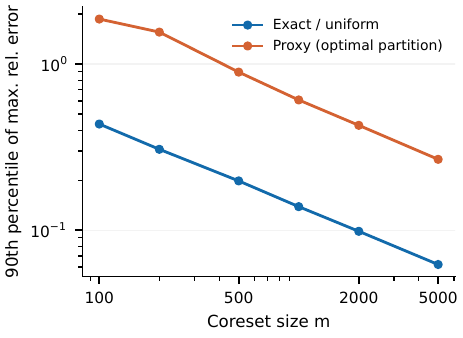}
        \caption{90th percentile of maximum relative cost error on the small simplex. Same settings as Figure \ref{fig:mean-simplex}; these are relative-error statistics, not multiplicative distortion.}
        \label{fig:p90-simplex}
    \end{minipage}
\end{figure}

The original witness is the center set
\[
S=\{x_1,\ldots,x_{k-1},0\}.
\]
The singleton vertices contribute zero cost to $S$, while the remaining $r$ vertices determine its input cost. We include this witness and 63 fixed randomly selected basis-plus-origin queries, using
\[
m\in\{100,200,500,1000,2000,5000\}
\]
and 10,000 draws per method and size. Because zero sampled cost gives infinite multiplicative distortion, Figures \ref{fig:mean-simplex} and \ref{fig:p90-simplex} plot the maximum absolute relative cost error over the 64 queries, rather than labelling it as distortion. At $m=2000$, mean maximum relative error is $0.0810$ for exact/uniform sampling and $0.3185$ for the proxy; 91.76\% of exact/uniform draws put every query within 10\% of its input cost, versus none for the proxy. At $m=100$, 36.61\% of proxy draws have an infinite-distortion witness. Sizes above $n=300$ illustrate a sampling-distribution effect but are not compression evidence.

\begin{figure}[!htbp]
    \centering
    \includegraphics[width=\linewidth]{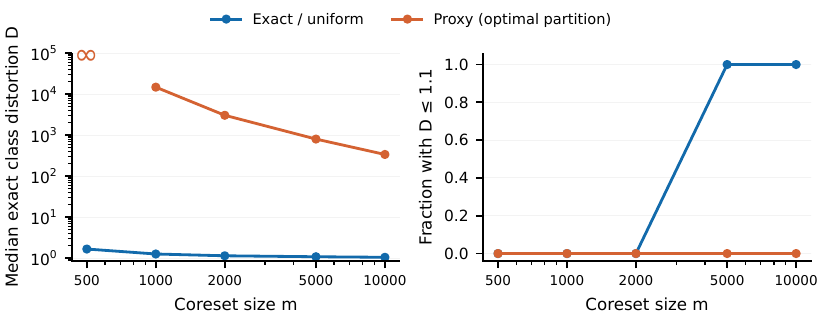}
    \caption{Compression-regime simplex, $n=30000$, $k=200$, and 2,000 draws per method and size. Left: median distortion, evaluated exactly over all basis-plus-origin queries; the infinity symbol records infinite median distortion. Right: fraction of draws with $D\leq1.1$. All tested sizes are smaller than the input.}
    \label{fig:large-simplex}
\end{figure}

To test a compression regime, we also use $n=30000$ and $k=200$, with 2,000 draws per method and size and
\[
m\in\{500,1000,2000,5000,10000\}.
\]
We evaluate exactly over the entire fixed class
\[
\mathcal{Q}_{\mathrm{basis}}=\{\{0\}\cup\{e_i:i\in A\}: A\subseteq[n],\ |A|=k-1\}.
\]
Every query has input cost $r=n-k+1$. If $W_i$ is the total sampled weight at vertex $i$, including $W_i=0$ for unsampled vertices, the minimum query cost is $\sum_i W_i$ minus the sum of the $k-1$ largest $W_i$, and the maximum subtracts the $k-1$ smallest instead. These order statistics give exact distortion over the whole class, without sampling center queries or storing a dense identity matrix. At $m=5000$, median distortion is $1.084$ for exact/uniform sampling and $804.033$ for the proxy; all 2,000 exact/uniform draws satisfy $D\leq1.1$, whereas none of the proxy draws do. At $m=500$, 99.95\% of proxy draws have infinite distortion. Figure \ref{fig:large-simplex} shows all tested sizes. This is an exact audit over the stated class, not a certificate over every Euclidean center set.

\FloatBarrier

\end{document}